\documentclass[doublecol]{epl2} 
\usepackage{color}

\title{Comment on ``Carnot efficiency at divergent power output'' (and additional discussion)}

\author{Y. Apertet}
\shortauthor{Y. Apertet}

\institute{                    
}

\abstract{In a recent Letter [EPL, 118 (2017) 40003], Polettini and Esposito claimed that it is theoretically possible for a thermodynamic machine to achieve Carnot efficiency at divergent power output through the use of infinitely-fast processes. It appears however that this assertion is misleading as it is not supported by their derivations as demonstrated below. In this Comment, we first show that there is a confusion regarding the notion of optimal efficiency. We then analyze the quantum dot engine described in Ref. [1] and demonstrate that Carnot efficiency is recovered only for vanishing output power.
Moreover, a discussion on the use of infinite thermodynamical forces to reach Carnot efficiency is also presented in the appendix.}

\begin{document}

\maketitle
\section{Introduction}
In a recent Letter \cite{Polettini2017}, Polettini and Esposito claimed that it is theoretically possible for a thermodynamic machine to achieve Carnot efficiency at divergent power output through the use of infinitely-fast processes. It appears however that this assertion is misleading as it is not supported by their derivations as demonstrated below. In this Comment, we first show that there is a confusion regarding the notion of optimal efficiency. We then analyze the quantum dot engine described in Ref.~\cite{Polettini2017} and demonstrate that Carnot efficiency is recovered only for vanishing output power.

\section{Main concerns}
Carnot efficiency sets an absolute limit on the efficiency with which heat energy can be turned into work. Yet, in Ref.~\cite{Polettini2017}, only work-to-work conversion is considered. As this kind of energy conversion is not limited by the Second law of Thermodynamics, the optimal associated efficiency is thus unity as stressed by Polettini and Esposito just below Eq.~(1) in Ref.~\cite{Polettini2017}. However, this notion of optimal efficiency is modified along Ref.~\cite{Polettini2017}: Indeed, an other optimal efficiency $\eta_{opt}$, different from unity, is introduced in Eq.~(14) of Ref.~\cite{Polettini2017}. It appears that this quantity is the actual limit of the efficiency when increasing tunneling rates. Infinitely increasing tunneling rates only is thus not sufficient to reach the genuine optimal efficiency, i.e., unity. This is our first main remark. 

\begin{figure}
	\centering
		\includegraphics[width=0.45\textwidth]{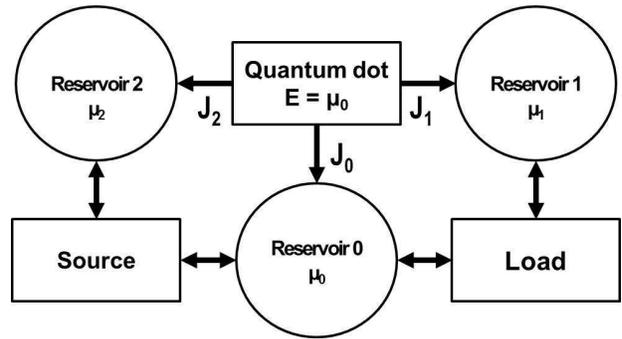}
	\caption{Schematic description of the QD presented in Ref.~\cite{Polettini2017} including source and load.}
	\label{fig:figure1}
\end{figure}
As highlighted by Eq.~(5) of Ref.~\cite{Polettini2017}, the optimization of the system is tightly associated with the choice of working conditions obtained through the applied forces. In order to properly consider the influence of these external forces, the schematic description of the quantum dot (QD) presented in Ref.~\cite{Polettini2017} is displayed on Fig.~\ref{fig:figure1} including source and load.
Under this form, the behavior of the system is easily understandable: Setting $\gamma_1 = \gamma_2 = \gamma \gg \gamma_0$ (keeping the same notations as in Ref.~\cite{Polettini2017}) amounts to favoring the exchanges between the QD and reservoirs 1 and 2 compared to the exchanges with reservoir 0. As $\gamma_0$ is set to 1, infinitely large tunneling rate $\gamma$ thus leads to isolate the QD from reservoir 0 since $J_0$ is then negligible compared to $J_1$ and $J_2$. The current conservation law imposes then that $J_1 = -J_2$. As both current are then proportional to each other, $\gamma \gg \gamma_0$ appears as the strong-coupling condition for this system. The efficiency then becomes $\eta = \eta_{opt} = (\mu_0 - \mu_1) /(\mu_0 - \mu_2)$. Note that this behavior associated in Ref.~\cite{Polettini2017} with infinitely-fast processes due to the infinitely large tunneling rate $\gamma$ could have been obtained setting instead $\gamma = 1$ and an infinitely-small tunneling rate $\gamma_0$ as only the ratio $\gamma / \gamma_0$ matters. While the latter setting leads to an identical behavior of the system, it presents the advantage to avoid an unphysical situation as discussed below. Infinitely-fast processes are thus not mandatory to reach the optimal efficiency $\eta_{opt}$. This is our second main remark.

While the authors of Ref.~\cite{Polettini2017} are interested in reaching ``Carnot efficiency'', it is surprising that they do not consider the working condition associated with maximum efficiency. Indeed, if the strong-coupling condition $\gamma_0/\gamma \rightarrow 0$ is satisfied, for given internal parameters of the system, namely tunneling rates and energy $E$, it is possible to approach Carnot efficiency setting the proper applied forces. Such efficiency optimization is obtained for the condition $F_1 \rightarrow F_2$. Note that such condition leads to vanishing output power. This is our third main remark. Imposing a different value for $F_1$ as it is the case in Ref.~\cite{Polettini2017} leads to lower efficiency. It can be seen in Fig.~3 of Ref.~\cite{Polettini2017} where the dotted curves are associated with efficiency maximization while the thick curves are obtained from Stochastic Thermodynamics for $F_1 - F_2 = 1$. So one may wonder if it is possible to reach Carnot efficiency without efficiency maximization, i.e., if it is still possible to get higher efficiency.


In Ref.~\cite{Polettini2017}, the authors suggest that it is possible to reach ``Carnot efficiency'' at finite power using infinite thermodynamic forces. However, as the thermodynamic forces $F_1$ and $F_2$ are increased, the charging rates $\omega_1^-$ and $\omega_2^-$ decrease. If the tunneling rate $\gamma$ is set as constant, an infinite increase of the forces would lead to infinitely small charging rates $\omega_1^-$ and $\omega_2^-$ since the occupation function at energy $E (= \mu_0)$ in each reservoir, given by Fermi distribution, would exponentially decrease toward zero. It leads to a situation where electrons could only flows from the QD to reservoirs 1 or 2 but not in the opposite direction. As a consequence, the QD is charged only by the reservoir 0. The source and the load both work as generators since currents $J_1$ and $J_2$ are then positive, as it is the case in the inset of Fig.(3) of Ref.~\cite{Polettini2017}, and the whole system behaves as a ``dud'' engine since the entire power coming from both load and source is dissipated as heat in the system. To avoid such useless working condition, it is possible to recover the strong-coupling condition, i.e., $J_0$ being negligible compared to $J_1$ and $J_2$, increasing the tunneling rate $\gamma$ in order to compensate the small value of the occupation function at energy $E$ in the expression of the charging rates. Hence, the higher the thermodynamic forces, the higher the value of tunneling rate needed to recover appropriate engine behavior is. This fact is clearly illustrated by Fig.(3) of Ref.~\cite{Polettini2017} since the minimum value of $\gamma$ necessary to get a positive efficiency increases with the ratio $\delta_1/\delta_2$, i.e., when $\eta_{opt} \rightarrow 1$. So, considering infinite forces imposes using infinite tunneling rate $\gamma$ to only envisage a proper behavior of the engine. Reaching maximum efficiency would require to go beyond this already infinite value.
Moreover, the perfect coupling between the QD and the reservoirs 1 and 2 associated with infinite tunneling rate $\gamma$ leads to some restrictions about the possible choice of working condition. Indeed, since the tunneling rate is inversely proportional to the tunneling resistance \cite{Devoret}, infinite tunneling rate thus implies vanishing tunneling resistance. Just as it is the case in a perfect electrical conductor, i.e., with vanishing electrical resistance, the potential difference at the edges of the system is then always zero whatever the electrical current is. This could also be understood using Heisenberg's uncertainty principle: An infinite coupling leads to an infinite broadening of the single energy level inside the QD. Consequently, assuming $\gamma \rightarrow \infty$ amounts setting $\mu_1 = \mu_2$ as if both reservoirs were connected through a perfect wire. So, even if the strong coupling condition could be achieved, the output power would then vanish.

Finally, it appears looking at Fig.~\ref{fig:figure1} that the QD used in Ref.~\cite{Polettini2017} cannot really be designated as an energy conversion machine. Indeed, as the input and output energies are of identical nature, there is no real conversion. The system might rather be seen as an energy valve letting more or less energy flowing from the source to the load. In the limit case $\gamma \gg \gamma_0$, the QD system even appears only as a dissipative component, dissipating a power $P = (\mu_1 - \mu_2)J_2$. Nonetheless, such dissipative components are useful on a fundamental level as they are mandatory to allow the connection between the source and the load, avoiding potential discontinuity between these two elements. Actually, it is possible to achieve the optimal efficiency for the energy transfer (rather than energy conversion) replacing the QD by a mere electrical resistance linking reservoir 1 and reservoir 2: In this case, one can increase the transfer efficiency toward unity considering the unphysical situation where thermodynamical forces $F_1$ and $F_2$ are infinitely large while the difference $(\mu_2 - \mu_1)$ remains finite. 

\section{Additional remark}

In Ref.~\cite{Polettini2017}, as currents $J_i$ are positive when QD is discharging, the conditions used by the authors lead to a positive current $J_2$, and so to a negative $J_1$ when strong coupling is ensured as then $J_2 = -J_1$. The forces $F_1$ and $F_2$ are negative as $\delta_1$ and $\delta_2$ are positive (according to the appendix of Ref.~\cite{Polettini2017}, $F_i = - \delta_i$ for the chosen conditions). According to Eq.~(2) of Ref.~\cite{Polettini2017}, both powers $P_1$ and $P_2$ should then be negative. However, the definitions of forces and fluxes used to characterize the QD engine are not consistent with the general description associated with linear regime. For the QD, the input and output power are actually given respectively by $P_2 = - F_2 J_2$ and $P_1 = F_1 J_1$ contrary to the definitions given in Eq.~(2) of Ref.~\cite{Polettini2017}. With these proper definitions, one notices that $P_1$ and $P_2$ are positive. The QD thus demonstrates the expected behavior, \textit{i.e.}, absorbing power from the source while releasing power into the load.

\section{Conclusion}

In Ref.~\cite{Polettini2017}, while they show for a QD that an optimal efficiency $\eta_{opt}$, different from Carnot efficiency, can be achieve with strong coupling assumption, \textit{i.e.}, $\gamma \gg \gamma_0$, the authors fails to clearly demonstrate that it is possible to actually reach Carnot efficiency at non vanishing power output. Furthermore, the approach used by Polettini and Esposito fails to reflect general properties of energy converters as they consider only systems transferring power rather than converting power.

\acknowledgments
I thank Dr. V. Holubec for fruitful discussion.

\section{Appendix: Reaching Carnot efficiency using infinite forces}

\begin{figure}
	\centering
		\includegraphics[width=0.45\textwidth]{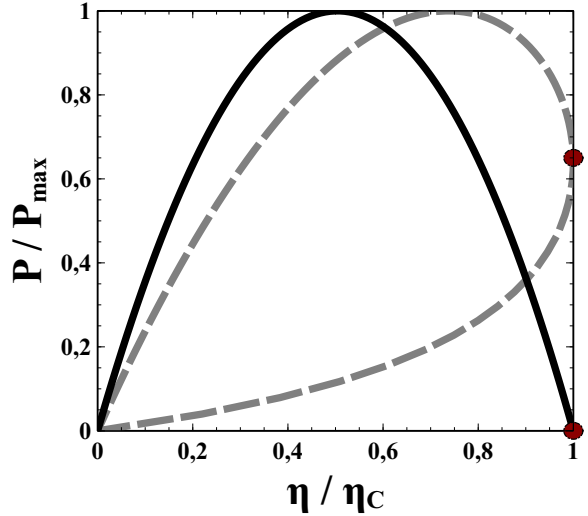}
	\caption{Normalized power $P / P_{\rm max}$ vs. relative efficiency $\eta / \eta_{\rm C}$, $\eta_{\rm C}$ being the Carnot efficiency, for a given autonomous heat engine. The curves are obtained for constant force $F_2$ by varying the force $F_1$ associated with the load. Full line : Typical power-efficiency characteristic for an heat engine working under strong-coupling condition. Dashed line: Typical power-efficiency characteristic for an heat engine working with broken time reversal symmetry as proposed by Benenti and coworkers in Ref.~\cite{Benenti2011} (from Ref.~\cite{Apertet2013}). The red dots correspond to working conditions associated with genuine Carnot efficiency.}
	\label{fig:courbe1}
\end{figure}

The approach presented by Polettini and Esposito in Ref.~\cite{Polettini2017} to reach Carnot efficiency at non vanishing power is based on the use of infinite force $F_1$ and $F_2$ rather than on infinitely-fast processes. To better illustrate this assertion, we consider the typical power-efficiency characteristic for an autonomous heat engine working under strong-coupling condition displayed on Fig.~\ref{fig:courbe1}. When the force $F_1$ associated with the load is increased from $0$ to $F_2$, the efficiency also increases while the output power $P$ first increases to reach its maximum value $P_{max}$ and then decreases to finally vanish when the maximum efficiency is reached. The two extreme working points of the curve, both associated with vanishing output power, correspond respectively to a short-circuit condition as $F_2 = 0$ and to an open-circuit condition as $J_2 = J_1 = 0$ since $F_1 = F_2$. This latter point is associated with Carnot efficiency only for strong-coupling condition: If it is not the case, the leakage between the two reservoirs prevent to obtain an ideal Carnot efficiency and the P-$\eta$ curve becomes a closed loop as the efficiency also vanishes for open-circuit condition.

In a recent Letter, Benenti and coworkers demonstrated that an heat engine working with broken time reversal symmetry, and limited only by the Second Law of Thermodynamics, may reach Carnot efficiency for finite output power. In this case, reaching this optimal efficiency no longer imposes strong-coupling condition to be verified. It is even quite the opposite as then a minimal parasitic heat flux is mandatory to guaranty the positivity of the entropy production during the energy conversion process \cite{Apertet2013}. The power-efficiency characteristic for such system is also displayed on Fig.~\ref{fig:courbe1}. It is clear from that curve that the efficiency maximization lead to Carnot efficiency and finite power at the same time.

\begin{figure}
	\centering
		\includegraphics[width=0.45\textwidth]{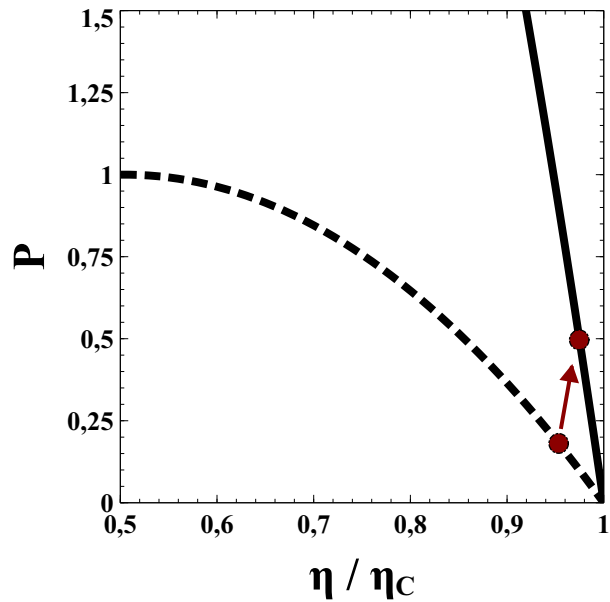}
	\caption{Power $P$ (in arbitrary unit) vs. relative efficiency $\eta / \eta_{\rm C}$ for two different values of the force $F_2$ associated with the source. The red dots correspond to working conditions associated with non vanishing power but close to Carnot efficiency.}
	\label{fig:courbe2}
\end{figure}

As already stressed in the main text, Polettini and Esposito \cite{Polettini2017} overlooked in their approach such efficiency maximization. On the contrary, they avoided it since they fix the difference between the forces $F_1$ and $F_2$ as $F_1 - F_2 = 1$. This choice might allow to set the working point for the system in the second half of the characteristic, i.e., $\eta > \eta(P_{max})$, and more particularly close to the limit working condition for which Carnot efficiency is reached as displayed on Fig.~\ref{fig:courbe2}. Once this working condition is set, it becomes quite easy to increase the efficiency while keeping finite output power: One only needs to increase the maximum efficiency $P_{max}$ for the system while keeping the strong-coupling condition. This operation will ``stretch'' the P-$\eta$ curve and, as the slope of the curve then increases, the efficiency also increases as illustrated on Fig.~\ref{fig:courbe2}. When the curve is infinitely stretched, i.e., for $P_{max} \rightarrow \infty$, the efficiency of the system tends to the Carnot efficiency even if the engine does not work under open-circuit condition. Since the chosen working point must be relatively close to this latter working condition (compared to the working condition associated with maximum power), the recent result of Holubec and Ryabov is verified: Carnot efficiency can be reached only at output power vanishingly small as compared to the maximum power $P_{max}$ attainable by the system \cite{Holubec2017}. The maximum power $P_{max}$ depends on both thermodynamical force $F_2$ and kinetic coefficients $L_{ij}$: It is then possible to use this strategy infinitely increasing the coupling parameter $L_{12}$ (= $L_{21}$) or the external force $F_2$. 

While this procedure is obviously valid from a mathematical viewpoint, it is more debatable from a physical viewpoint: Indeed, even if the efficiency tends to Carnot efficiency, for a given engine, it is always possible to obtain a \emph{higher} efficiency by approaching the open-circuit condition, unfortunately associated with vanishing output power. So, it remains unclear if this method can truly be viewed as a way to challenge the common wisdom that efficiency can only be optimal in the limit of infinitely-slow processes without invoking broken time reversal symmetry.
In any case, the example of the QD system presented by Polettini and Esposito in Ref.~\cite{Polettini2017} demonstrates that even if this strategy seems to be quite simple, its application can be tricky. Indeed, as discussed in the main text of the Comment, setting the proper working condition close to the open-circuit condition and maintaining strong-coupling inside the engine while external forces are increased can be complex, especially if one adds another constraint, e.g., a constant difference between $F_1$ and $F_2$ as it is the case in Ref.~\cite{Polettini2017}. Furthermore, considering infinite thermodynamical forces may involve some fundamental limitations (e.g., level broadening in a QD), not to mention the practical limitations, preventing the engine to work the way it was meant to work. 

From the above observations, we can now understand why infinite transition rates, associated with infinitely-fast processes in Ref.~\cite{Polettini2017}, are not mandatory in the general case: The need for an infinite rate $\gamma$ stems only from the need to ensure the strong-coupling condition for the QD whatever the values of the forces $F_1$ and $F_2$ are. Since strong-coupling condition amounts to neglecting the connection between the QD and reservoir $0$, the system with strong-coupling can be viewed as a very complex voltage divider, but still a voltage divider. If one considers instead a simpler voltage divider, i.e., an ideal resistor between reservoirs $1$ and $2$, it is straightforward to show that efficiency tends to unity when forces are infinitely increased while keeping a small constant difference between the two forces. Moreover, even if the electrical output power becomes infinite, the electrical current flowing through the resistor remains finite in this case.


\begin{thebibliography}{12}
\bibitem{Polettini2017}
  \Name{Polettini M. \and Esposito M.}
  \REVIEW{EPL}{118}{2017}{40003}.

\bibitem{Devoret}
\Name{Grabert H. \and Devoret M. H. (Eds.)}
\Book{Single Charge Tunneling: Coulomb Blockade Phenomena In Nanostructures}
\Publ{Springer Science \& Business Media}
\Year{1992}.



\bibitem{Benenti2011}
	\Name{Benenti G., Saito K., \and Casati G.}
	\REVIEW{Phys. Rev. Lett.}{106}{2011}{230602}.

\bibitem{Apertet2013}
	\Name{Apertet Y., Ouerdane H., Goupil C., \and Lecoeur Ph.}
	\REVIEW{Phys. Rev. E}{88}{2013}{022137}.

\bibitem{Holubec2017}
  \Name{Holubec V. \and Ryabov A.}
  \REVIEW{Phys. Rev. E}{96}{2017}{062107}.


\end{thebibliography}
\end{document}